\documentstyle[psfig]{article}
\newlength{\TWidth}
\newcounter{HeadingLevel}

\newcommand{\STRUT}{\rule{0mm}{1mm}}

\newcommand{\Title}[1]{
\vspace*{24.090pt}\indent\parbox[t]{\TWidth}{#1}\\*[24.090pt]}

\newcommand{\Author}[1]{\STRUT\hfill\parbox[t]{\TWidth}{#1}\\}
\newcommand{\Affiliation}[1]{\STRUT\hfill\parbox[t]{\TWidth}{#1}\\*[1em]}

\newcommand{\Heading}[1]{
\ifthenelse{\theHeadingLevel = 1}{\vspace{-1em}}{\vspace{1em}}
\setcounter{HeadingLevel}{0}
\par\noindent\underline{#1}\\
}

\begin{document}

\hyphenation{clas-si-fy}
\hyphenation{Par-tha-sa-ra-thy}
\Title{THE VELOCITY GRADIENT IN THE PSEUDO-PHOTOSPHERE \\
OF THE PECULIAR SUPERGIANT HD 101584}
\Author{ERIC J. BAKKER}
\Affiliation{SRON Laboratory for Space Research Utrecht \hfill \break
Sorbonnelaan 2 \hfill \break
3584 CA UTRECHT \hfill \break
The Netherlands}

\begin{abstract}
In this paper preliminary results are presented based on a study of
the low and high resolution ultraviolet spectrum
of the peculiar supergiant (post-AGB star) HD 101584.
By a comparison of the low resolution spectrum ($1200-3200 {\rm~\AA}$) with
standard stars, the star is classified as an A7I, indicating an effective
temperature of 8150 K, where literature quotes spectral type F0I.
The Doppler shift of the FeII absorption lines in the high resolution spectrum
($2500-3000{\rm~\AA}$) show a relation with the line optical depth. This
suggests an expanding accelerating wind, c.q. pseudo-photosphere.
The    relation is extended by a factor $10^{5}$ in optical depth by
using available data from optical HeI and NI lines.
The relation suggests that the radial heliocentric velocity of the star
is at least $54.5 {\rm~km~s^{-1}}$.
{}From the ${\rm H} \alpha $ line a
velocity of $ 96 {\rm ~km~s^{-1}}$ is measured for the terminal velocity of
the wind.
\end{abstract}

\section{Introduction}

The star HD 101584 ($ {\rm b} = 6 ^{o} $)
is classified as a 7.01 visual magnitude F0Iape
with $ (\rm{B - V} )= +0.39 $
(Hoffleit 1983), indicating an
effective temperature of $7700 {\rm~K}$ , $ \log {\rm~ g}  =1.7 $ , and
thus $ (\rm{B - V})_{0}  = 0.17 $
(Landolt-B\"{o}rnstein 1982). Far- and near-infrared
photometry reveals a strong infrared source at the position of the star
(Humphreys \& Ney 1974; Parthasarathy \& Pottasch 1986).
Molecular line observations show bipolar outflow for the OH maser
(te Lintel Hekkert et al. 1992)
and a very complex structure
for the $ {\rm CO} ( {\rm J}= 1 \rightarrow 0 ) $ transition
(Trams et al. 1990; Loup et al. 1990; van der Veen et al. 1992).
The molecular line emission in OH and CO is normally
discussed in terms of evolved stars, and fits the idea that HD 101584 is
a post-AGB star.

The first to classify the star HD 101584 as a post-AGB star were
Parthasarathy \& Pottasch (1986).  Their conclusion was based on the
strong infrared excess of the star
which seems to be due to a large amount of dust around
the star. An extensive study by Trams et al.
(1991) shows the resemblance of the infrared excess
of HD 101584 with other known post-AGB stars.
It is
now reasonably well established that HD 101584 is a post-AGB star.

\section{The ultraviolet spectrum}

\subsection{The low resolution IUE spectrum}

\begin{figure*}
\centerline{\hbox{\psfig{figure=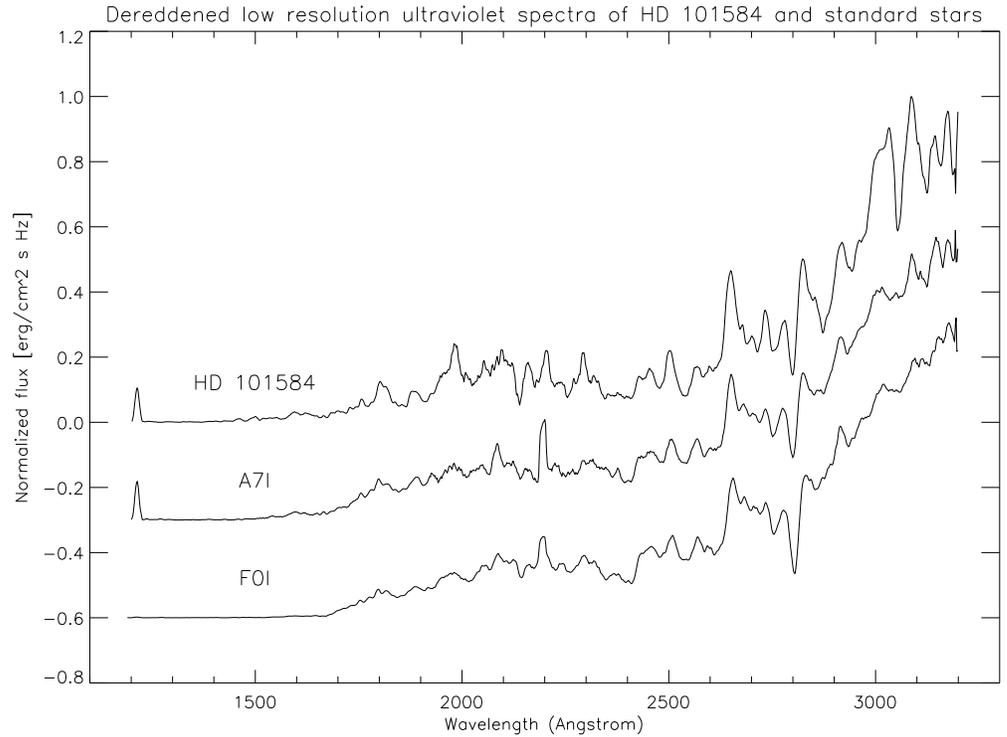,height=10cm}}}
\caption{The low resolution IUE spectrum of HD 101584 (upper),
         a standard A7I star (middle), and a standard F0I star (lower).
         The spectra are smoothed over $ 12.6 {\rm~\AA}$, normalized,
         and dereddened}
\label{lhlfig-lores}
\end{figure*}

The low resolution IUE spectra  ($ 9 {\rm AA} $)
of HD 101584, a A7I and a F0I standard
star are shown in  figure~\ref{lhlfig-lores}.
Before fitting the ultraviolet energy distribution
to a standard star the spectrum was smoothed over $ 12.6 {\rm~\AA} $
and corrected for interstellar and circumstellar extinction
(Mathis 1990)
using a colour excess, ${\rm E(B-V)}=0.27$, derived for spectral type A7I.

The low resolution ultraviolet spectrum was best fitted to the
standard A7I star, HD14873. An even better fit is possible
if also an A6I star would be available in the reference atlas
(Heck et al. 1984). The spectrum of HD 101584 shows no flux lower then
$1400 {\rm ~\AA}$ and can therefore not be fitted with
a spectrum of a star of spectral type A5I or hotter.

By comparing the spectrum of HD 101584 with the standard
F0I star, $ \alpha {\rm~Lep} $, an excess of flux for
the program star between $ 1400 \mbox{ and } 1700 {\rm ~\AA} $ indicates
a higher temperature of the star. The slope of the continuum of the
spectrum confirms the supergiant nature of the star. The data on the
spectrum of HD 101584 and on the two reference stars is in
table~\ref{lhltab-lowspec}.

\begin{table*}
\caption[]{Data on the low resolution spectra}
\label{lhltab-lowspec}
\begin{flushleft}
\begin{tabular}{lllll}
\hline
HD number&Name              &Spectral&E(B-V)&Normalized on               \\
         &                  &type    &      &[${\rm erg~cm^{-2}~s~Hz}$]  \\
\hline
HD 101584&                  &A7I     &0.27  &$4.1~10^{-23}$              \\
HD  36673&$\alpha {\rm~Lep}$&F0I     &0.04  &$6.0~10^{-22}$              \\
HD 148743&                  &A7I     &0.25  &$2.8~10^{-23}$              \\
\hline
\end{tabular}
\end{flushleft}
\end{table*}

\begin{table*}
\caption[]{Stellar parameters for HD 101584 based on
literature and on the low
resolution ultraviolet IUE spectra studied in this work}
\begin{flushleft}
\begin{tabular}{llllll}
\hline
Work &Spectral & Effective   & $ \log {\rm~ g} $ & $ ({\rm B - V} ) _{0} $ &
$ {\rm E ( B - V )} $ \\
     &Type     & Temperature &            &
&
                              \\
\hline
Hoffleit 1983 & F0I      & 7700        & 1.7       & +0.17
        &
0.22 \\
This study    & A7I      & 8150        & 1.9     & +0.12
      &
0.27 \\
\hline
\end{tabular}
\end{flushleft}
\par
${\rm T_{eff}}$, $\log {\rm~g}$, and ${\rm (B-V)_{o}}$ from
Landolt-B\"{o}rnstein (1982) and are based on the given spectral type.
\end{table*}

\subsection{the high resolution IUE spectrum}

In an extensive study of the high resolution
($0.3{\rm~ \AA} $)
IUE spectrum of HD 101584 a large number of absorption features
between $ 2500 $ and $ 3000{\rm~ \AA} $ has been identified (Bakker 1994).
The main conclusions from this work are that the spectrum of HD 101584
has in principle the same absorption features as the F0 supergiant,
$ \alpha {\rm~Lep} $, but the lines are intrinsically broader
and the lines are asymmetric in shape.
This study limits itself to the
measured radial velocities of the absorption features as
derived  from the measured Doppler shift of the core of the
absorption profile.

\begin{figure*}
\centerline{\hbox{\psfig{figure=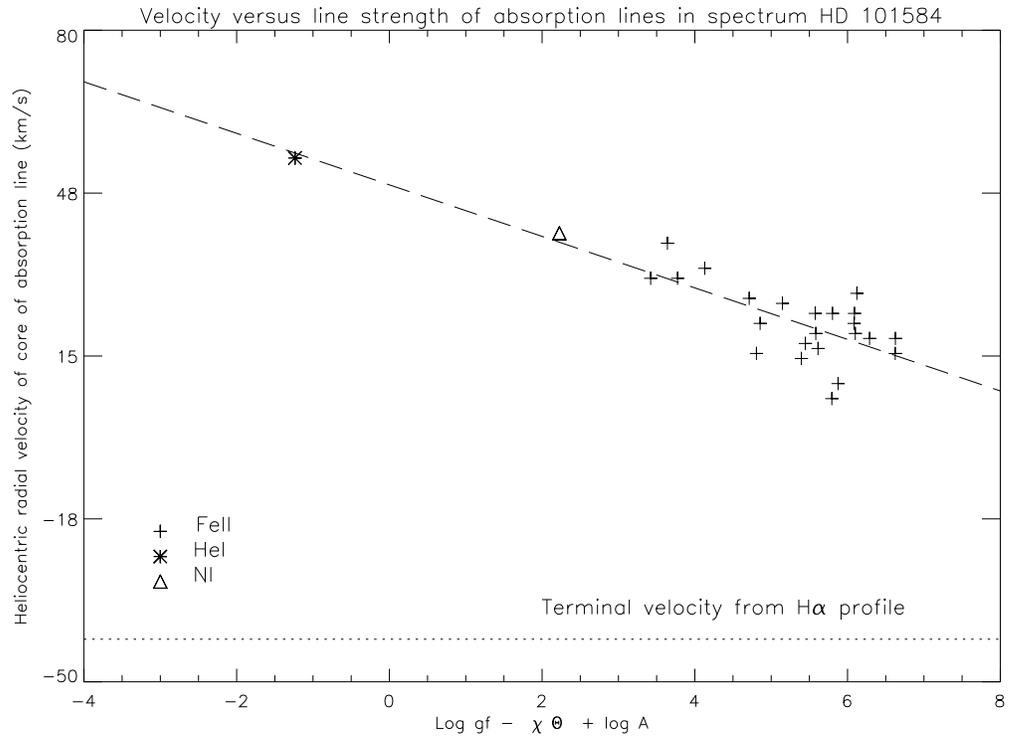,height=10cm}}}
\caption{Relation between the strength of an absorption line
         and the radial velocity measured for that line, indicating that
         we are looking at an accelerating expanding wind.
         The dashed line is a first order fit based on the FeII lines.
         The HeI and NI line seem to fit this relation exceptionally well.
         Where $\theta$ is $5040/{\rm T_{eff}}$, $\chi$ is the
         excitation potential of the lower level of the transition, and
         A is the abundance relative to H by number}
\label{lhlfig-rel}
\end{figure*}

In probing the photosphere of a star the line optical depth
is a measure for the depth in the photosphere seen. The
stronger the optical depth $ \tau $ the more outside layers
of the photosphere will be probed. The range in optical depth from
the ultraviolet spectrum is limited to about a
factor $ 10^{4} $. By incorporating some of the available optical
data, which are from  much weaker lines, the deeper layers of the
photosphere can be probed as well and the relation can
be extended over a much wider range of $\tau $
by a factor of $10^{5}$ to
$ 10^{9} $. There are however two assumptions in making the
relation  valid for a larger range of $ \tau $.
The first is solar abundance ratios,
and that NI, HeI, and FeII are the dominant ionization stages of these
elements in the stellar wind (or pseudo-photosphere). The second is that the
effective temperature derived from the energy distribution in the ultraviolet
represents the real temperature of the gas. If the first
assumption is violated, the separate elements will shift horizontally
in the fig.~\ref{lhlfig-rel}. If the second assumption is violated,
the relation within
FeII will change, and there will occur a small horizontal shift
between the different elements.
The data on FeII, NI, HeI and ${\rm H} \alpha $ will be published in a
separate paper which is in preparation (Bakker 1995).

Figure~\ref{lhlfig-rel} shows the relation between the
logarithm of the strength  of an absorption
line and the heliocentric radial velocity measured for that line.
The crosses are the FeII lines from the ultraviolet spectrum, the
triangle the optical nitrogen line
($8680.24 {\rm~\AA}$), and the asterisk the optical helium line
($5875.618 {\rm~ \AA}$).

\begin{equation}
\label{lhleq-rel}
V_{helio}({\rm FeII}) = 49.1 -5.1 \times \left[
\log {\rm~ N} + \log {\rm ~gf} - \frac{5040 \chi }{{\rm T_{eff}}}  \right]
\left[ km~s^{-1} \right]
\end{equation}

\begin{table*}
\caption[]{Main parameters of the chemical elements used in probing the
pseudo-photosphere}
\label{lhltab-par}
\begin{flushleft}
\begin{tabular}{ll|lll}
\hline
Element   & Solar  Abundance    & Ion       & Ionization  & Excitation   \\
          & $ \log {\rm~ N} $   &           & Energy (eV) & Energy (eV)  \\
\hline
He        & 10.93               & HeI       & 24.587      & 20.87         \\
N         &  7.96               & NI        & 14.534      & 10.29         \\
Fe        &  7.60               & FeII      & 16.16       &
 $ 7.870 \rightarrow  4.48 $ \\
\hline
\end{tabular}
\end{flushleft}
\par
Solar Abundance, and ionization energies are from Allen (1985).
\end{table*}

A first order approximation of the relation for the FeII line is
given by eq.~\ref{lhleq-rel}. The reason that the HeI and NI line are not
used for this first order approximation is that is seems logical to
assume a constant velocity for weak lines. These lines are formed in
the deeper layers of the photosphere and are therefore least
affected by the unknown force which accelerates the pseudo-photosphere
outwards. The dashed line in fig.~\ref{lhlfig-rel} shows
the first order approximation based on only the FeII lines.
It is surprising to see that the HeI and NI line fit this relation
almost perfectly.

\section{Terminal velocity of the stellar wind}

An upper limit on the maximum out streaming velocity (a lower limit
on the blue shift of a line)  is given by the absorption
part of the $ {\rm H} \alpha $ line profile.
The edge of the
$ {\rm H} \alpha $ line profile is pretty steep,  implying
that hydrogen column density at that velocity does not slowly
decrease due to expansion (and thus dilution) of the gas,
but rather that the edge in the profile is due to hydrogen
at the terminal velocity of the stellar wind.
The maximum out streaming velocity is $120{\rm~km~s^{-1}}$, and
the out-streaming velocity derived from the Doppler shift of the
core of the absorption profile is $96{\rm~km~s^{-1}}$.
The latter velocity is the terminal
velocity of the wind. The first is the net velocity maximum due to
turbulent motion in the wind and its expansion.
If we assume that the velocity of the star
is best represented by the
velocity of the HeI line, a minimum heliocentric velocity of
$ -41 {\rm ~km~s^{-1}}$ is expected.
The dotted line in fig.~\ref{lhlfig-rel} represents the terminal velocity
of the wind as determined from the ${\rm H} \alpha $ profile.

\section{Discussion}

The data in the literature concerning  the radial velocity of HD 101584 is
very confusing. It's variations are not well understood and
is a fruitful base for wild speculations.
{}From this study a little light is shed on these variations.
The relation as shown in figure~\ref{lhlfig-rel} reveals that
variations in radial velocity within one spectrum (FeII lines) can be
understood
in terms of an expanding accelerating photosphere or wind.
In the following
discussion this line absorbing region will be called pseudo-photosphere.
This relation can
be extended to lower optical depth by incorporating the helium
($5875.618 {\rm~\AA}$)  and nitrogen ($8680.24 {\rm~\AA}$)
optical absorption lines. That this relation holds for helium suggests that
this line is from the same star and not from a yet unseen hot companion
star. To produce a helium absorption line the star
has to be much hotter than spectral type A7I, probably even a B-type star.
This contradicts the fact that the low resolution ultraviolet spectrum
is best fitted with a A7I reference star

Although the relation does not seem to go to a constant velocity
for weaker lines (HeI and NI) is seems reasonable to
assume that the velocity for the HeI line is the stellar velocity.
By monitoring the velocity of the HeI absorption line it
should be possible to make a statement about the binary nature of the
star.
Radial velocity measurements of stronger lines do not only have
a contribution of the radial velocity of the star due to
binarity (if this would be the case), but will also have
a contribution from the pseudo-photosphere.
These two contributions will be very hard to disentangle.
A maximum velocity of the out streaming wind as determined from
the ${\rm H} \alpha $ profile is $96 {\rm~km~s^{-1}}$. This means that
velocities of absorption lines are to be expected in the range between
$-41 {\rm~km~s^{-1}}$ and $54.5 {\rm~km~s^{-1}}$.

Table~\ref{lhltab-dat} summarizes the stellar parameters
of HD 101584 as determined in this study from ultraviolet spectra.

\begin{table*}
\caption{Improved data on HD 101584}
\label{lhltab-dat}
\begin{flushleft}
\begin{tabular}{ll}
\hline
Ultraviolet spectral type          & A7I   \\
Effective temperature              & 8150 K\\
$ {\rm E ( B - V )} $              & 0.27  \\
Heliocentric velocity of star      & $ 54.5 {\rm~ km~s^{-1}}$\\
Terminal velocity of wind          & $ 96   {\rm~ km~s^{-1}}$\\
                                   &        \\
Maximum helio. velocity expected   & $ 54.5 {\rm ~km~s^{-1}}$ \\
Minimum helio. velocity expected   & $ -41  {\rm ~km~s^{-1}}$ \\
\hline
\end{tabular}
\end{flushleft}
\end{table*}

{\it Acknowledgements}
The author would like to thank Dr. Norman Trams for making the
HIRES spectrum of HD 101584 available for this work. Dr. Christoffel
Waelkens was so kind to make
the optical data of HeI, NI, and $ {\rm H} \alpha $ lines available.
Dr. Rens Waters and Prof. Dr. Henny Lamers made a large contribution
to this work by having many discussion with the author.
The author was supported by grant no. 782-371-040 by ASTRON,
which receives funds from the Netherlands Organization for
the Advancement of Pure Research (NWO).
This research has made use of the Simbad database, operated at
CDS, Strasbourg, France.

\end{document}